\documentclass[twocolumn,showpacs]{revtex4}
\usepackage{epstopdf}
\usepackage{graphicx}
\usepackage{dcolumn}
\usepackage{bm}

\def\gappeq{\mathrel{ \rlap{\raise.5ex\hbox{$>$}}
                      {\lower.5ex\hbox{$\sim$}}  } }
\def\lappeq{\mathrel{ \rlap{\raise.5ex\hbox{$<$}}
                      {\lower.5ex\hbox{$\sim$}}  } }

\begin{document}

\preprint{PRA}

\title{Dynamical Instability of a Rotating Dipolar Bose-Einstein Condensate.}

\author{R.M.W. van Bijnen$^{1,2}$, D.H.J. O'Dell$^3$, N.G. Parker$^{1}$ and A.M. Martin$^1$}

\address{$^{1}$ School of Physics, University of Melbourne, Parkville,
Victoria 3010, Australia. \\ $^{2}$ Eindhoven University of
Technology, PO Box 513, 5600 MB Eindhoven, The Netherlands. \\$^{3}$
Centre for Cold Matter, Imperial College, Prince Consort Road,
London, SW7 2BW, United Kingdom.}

\date{\today}

\begin{abstract}
We calculate the hydrodynamic solutions for a dilute Bose-Einstein
condensate with long-range dipolar interactions in a rotating,
elliptical harmonic trap, and analyse their dynamical stability. The
static solutions and their regimes of instability vary non-trivially
on the strength of the dipolar interactions. We comprehensively map
out this behaviour, and in particular examine the experimental
routes towards unstable dynamics, which, in analogy to conventional
condensates, may lead to vortex lattice formation. Furthermore, we
analyse the centre of mass and breathing modes of a rotating dipolar
condensate.
\end{abstract}

\pacs{03.75.Kk, 34.20.Cf, 47.20.-k} \maketitle In recent years a
considerable amount of experimental \cite{Expt1,Expt2} and
theoretical
\cite{Recati,Sinha,Rapid,Madison,Makoto,Kenichi,Emil,Lobo,Nick1}
work has been carried out on dilute Bose-Einstein condensates (BECs)
in rotating anisotropic traps. Where short-range interactions
dominate, a vortex lattice forms when the rotational frequency
($\Omega$) of the system is $\approx 0.7 \omega_{\perp}$ (where
$\omega_{\perp}$ is the trapping frequency perpendicular to the axis
of rotation). Insight into the mechanism of vortex formation can be
gained by noting that $0.7 \omega_{\perp}$ closely coincides with
the frequency at which certain hydrodynamic surface excitations
become unstable \cite{Sinha,Rapid}. Through comparison with
experimental results \cite{Expt1,Expt2,Madison} and numerical
solutions of the Gross-Pitaevskii equation (GPE)
\cite{Rapid,Lobo,Nick1} such instability has been directly related
to the formation of a vortex lattice.


The above results apply to {\it conventional} BECs composed of atoms
of mass $m$ with short-range s-wave interactions, parameterised via
$g=4 \pi \hbar^2 a/m$, where $a$ is the s-wave scattering length.
However, a recent experiment has formed a BEC of chromium atoms with
dipolar interactions \cite{expt}. This opens the door to
experimentally study the effect of dipolar interactions in BECs.
Parallel theoretical work, using a modified GPE, has studied the
effect of such long-range interactions on the ground state vortex
lattice solutions \cite{lattice1,lattice2,lattice3}. However, the
route to generating such states has not been explored. For this
purpose we solve the hydrodynamic equations of motion for a dipolar
BEC in rotating anisotropic harmonic traps. We show that the
solutions depend on both the strength of the dipolar interactions,
$\varepsilon_{dd}$, and the aspect ratio of the trap,
$\gamma=\omega_z/\omega_{\perp}$, in stark contrast to conventional
BECs where they are independent of both the strength of the
interactions and $\gamma$ \cite{Recati,Sinha}. In addition we
evaluate the dynamical stability of our solutions, showing that the
region of $\Omega$ for which the solutions are stable can be
controlled via both $\varepsilon_{dd}$ and $\gamma$. By analogy to
conventional BECs \cite{Recati,Sinha,Rapid}, one may expect these
instabilites to result in vortex lattice formation in dipolar BECs.


Consider a BEC with long-range dipole-dipole interactions. The
potential between dipoles, separated by $\underline{r}$ and
aligned by an external electric or magnetic field along a unit vector
$\hat{e}$ is given by, in the notation of Ref. \cite{Duncnote},
\begin{eqnarray}
U_{dd}(\underline{r})=\frac{C_{dd}}{4 \pi} \hat{e}_i\hat{e}_j
\frac{\left(\delta_{ij}-3\hat{r}_i
\hat{r}_j\right)}{r^3}. \label{dipole_potential}
\end{eqnarray}
For low energy scattering of two
atoms with dipoles induced by a static electric
field $\underline{E}=E \hat{e}$, the coupling constant
$C_{dd}=E^2 \alpha^2/\epsilon_0$ \cite{You,Yi}. Alternatively, if
the atoms have permanent magnetic
dipoles, $d_m$, aligned in an external magnetic field
$\underline{B}=B \hat{e}$, one has $C_{dd}= \mu_0
d_m^2$ \cite{goral00}. Denoting $\rho$ as the condensate density,  the dipolar interactions give rise to a mean-field potential
\begin{equation}
\Phi_{dd} (\underline{r}) = \int d^3 r^{\prime} U_{dd} \left(
\underline{r}-\underline{r}^{\prime} \right)
\rho\left(\underline{r}^{\prime} \right) \label{dipole_a}
\end{equation}
which can be included in a generalized GPE
\cite{goral00,Yi,santos00} for the BEC. In the Thomas-Fermi (TF)
regime \cite{note1} the GPE describing a static dipolar BEC in a
harmonic
 trapping potential
$V(\underline{r})=m(\omega_x^2 x^2+\omega_y^2 y^2 +\omega_z^2
z^2)/2$ is
\begin{eqnarray}
\mu=\frac{m}{2}\left(\omega_x^2 x^2 +\omega_y^2 y^2 + \omega_z^2
z^2 \right) + g \rho(\underline{r}) + \Phi_{dd} (\underline{r}),
\label{mu_static}
\end{eqnarray}
where $\mu$ is the chemical potential. For ease of calculation the dipolar potential
$\Phi_{dd}(\underline{r}) $ can be expressed in terms of a fictitious
`electrostatic' potential $\phi(\underline{r})$ \cite{Duncan1}
\begin{eqnarray}
\Phi_{dd} (\underline{r}) & =& -3g \varepsilon_{dd} \hat{e}_i
\hat{e}_j \left(\nabla_i \nabla_j \phi(\underline{r})
+\frac{\delta_{ij}}{3} \rho(\underline{r}) \right) \label{eq:Phidd}
\label{dipole}
\end{eqnarray}
where $\quad \phi(\underline{r})  = \int d^3 r^{\prime}
\rho(\underline{r}^{\prime})/(4 \pi \left|
\underline{r}-\underline{r}^{\prime} \right|)$
and $\varepsilon_{dd}=C_{dd}/3g$ parameterizes the relative strength
of the dipolar and s-wave interactions. Self-consistent solutions of
Eq.\ (\ref{mu_static}) for $\rho(\underline{r})$,
$\phi(\underline{r})$ and hence $\Phi_{dd}(\underline{r})$ can be
found for any general parabolic trap, see Appendix A of Ref.
\cite{Duncan1} .

We consider atoms trapped in a harmonic potential rotating at a
frequency $\Omega$ about the $z$-axis. In the mean field
approximation the evolution of the condensate field,
$\psi(\underline{r},t)$, is described by the time-dependent GPE.
Writing the condensate field in terms of a density
$\rho(\underline{r})$ and a phase $S(\underline{r})$ and neglecting
the quantum pressure we obtain the conventional superfluid
hydrodynamic equations
\begin{eqnarray}
\frac{\partial \rho}{\partial t}+\nabla \cdot \left[\rho\left(\underline{v}-\underline{\Omega} \times \underline{r}\right)\right]=0
\label{continuity}
\end{eqnarray}
\begin{eqnarray}
\frac{\partial \underline{v}}{\partial t} + \nabla
\left(\frac{\underline{v} \cdot \underline{v}}{2} +
\frac{V}{m}+\frac{g \rho}{m} + \frac{\Phi_{dd}}{m} -\underline{v}
\cdot \left[ \underline{\Omega} \times
\underline{r}\right]\right)= 0, \label{hydro}
\end{eqnarray}
where $\underline{v}=(\hbar/m) \nabla S$ is the fluid velocity
field, in the laboratory frame, expressed in terms of the
coordinates in the rotating frame. The stationary solutions of
Eqs.~(\ref{continuity}) and~(\ref{hydro}) are obtained by imposing
the conditions $\partial \rho /\partial t =0$ and $\partial
\underline{v} / \partial t=0$. We look for solutions of the form
\cite{Recati,Sinha} $\underline{v}=\alpha (y \hat{i} +x \hat{j})$,
 where $\alpha$ is to be determined. We can combine this with Eq.~(\ref{hydro}) to obtain
\begin{eqnarray}
\mu=\frac{m}{2} \left(\tilde{\omega}_x^2 x^2 +\tilde{\omega}_y^2 y^2 +\omega_z^2 z^2 \right)+g\rho(\underline{r})+\Phi_{dd}(\underline{r})
\label{mu}
\end{eqnarray}
where $\tilde{\omega}_x^2=\omega_x^2+\alpha^2-2\alpha \Omega$ and
$\tilde{\omega}_y^2=\omega_y^2+\alpha^2+2\alpha \Omega$. The form
of Eq.~(\ref{mu}) is identical to Eq.~(\ref{mu_static}). Hence we
can use the methodology presented in Ref. \cite{Duncan1} to
calculate $\Phi_{dd}(\underline{r})$. An exact solution of Eq.\ (\ref{mu}) is given
by
\begin{eqnarray}
\rho=n_0\left(1-\frac{x^2}{R_x^2}-\frac{y^2}{R_y^2}-\frac{z^2}{R_z^2}\right) \,\,\,\, {\rm for} \,\,\, \rho \ge 0
\label{TF}
\end{eqnarray}
where $n_0$ is the central density which is given by normalization
to be $n_0=15N/(8 \pi R_x R_y R_z)$. Following the results
presented in Appendix A of Ref. \cite{Duncan1} the dipole
potential for a polarizing field aligned along the $z$-axis is
\begin{eqnarray}
\frac{\Phi_{dd}}{3g\varepsilon_{dd}}=\frac{n_0 \kappa_x
\kappa_y}{2}\left[\beta_0-\frac{x^2\beta_x+y^2\beta_y+3z^2\beta_z}{R_z^2}\right]
- \frac{\rho}{3}
\end{eqnarray}
where
\begin{eqnarray}
\beta_k=\int_0^{\infty} \frac {d \sigma}{\left(1+\sigma \right)\left(\kappa_k^2 + \sigma\right)\sqrt{\left(\kappa_x^2+\sigma\right)\left(\kappa_y^2+\sigma\right)\left(1+\sigma\right)}} \nonumber \\
\end{eqnarray}
with $k=x,y,z$, $\kappa_k=\frac{R_k}{R_z}$ and
\begin{eqnarray}
\beta_0=\int_0^{\infty} \frac{d \sigma}{\left(1+\sigma \right)\sqrt{\left(\kappa_x^2+\sigma\right)\left(\kappa_y^2+\sigma\right)\left(1+\sigma\right)}}.
\end{eqnarray}
Thus we can rearrange Eq.~(\ref{mu}) to obtain the  density
\begin{eqnarray}
\rho=\frac{\mu-3g\varepsilon_{dd}n_0\kappa_x\kappa_y\beta_0-\frac{m}{2}\left(\tilde{\tilde{\omega}}_xx^2+\tilde{\tilde{\omega}}_yy^2+\tilde{\tilde{\omega}}_zz^2\right)}{g\left(1-\varepsilon_{dd}\right)}
\label{rho}
\end{eqnarray}
where
$\tilde{\tilde{\omega}}_{x\{y\}}=\tilde{\omega}_{x\{y\}}^2-3\varepsilon_{dd}\kappa_x\kappa_y\beta_{x\{y\}}\omega_z^2/(2\zeta)$,
$\tilde{\tilde{\omega}}_{z}=\omega_z^2\left(1-9\varepsilon_{dd}\kappa_x\kappa_y\beta_{z}\right)/(2\zeta)$
and $\zeta=1-\varepsilon_{dd}\left[1-\frac{9 \kappa_x
\kappa_y}{2}\beta_z\right]$. Comparing the $x^2$, $y^2$ and $z^2$
terms in Eq.~(\ref{TF}) and Eq.~(\ref{rho}) we find the three
self-consistency relations:
\begin{eqnarray}
\kappa_{x\{y\}}^2=\left(\frac{\omega_z}{\tilde{\omega}_{x\{y\}}}\right)^2
\frac{1+\varepsilon_{dd}\left(\frac{3}{2}\kappa_{x\{y\}}^3\kappa_{y\{x\}}
\beta_{x\{y\}}-1\right)}{\zeta} \label{kxky}
\end{eqnarray}
and $R_z^2=\frac{2gn_0}{m\omega_z^2}\zeta$. Now using
Eq.~(\ref{continuity}) in conjunction with Eq.~(\ref{rho}) we find
the following expression for the stationary solutions to
Eq.~(\ref{continuity})
\begin{eqnarray}
0&=&\left(\alpha+\Omega\right)\left(\tilde{\omega}_x^{2}-\frac{3}{2}\varepsilon_{dd} \frac{\omega_x^2\kappa_x\kappa_y \gamma^{2}}{\zeta} \beta_x\right) \nonumber \\
&+&\left(\alpha-\Omega\right)\left(\tilde{\omega}_y^{
2}-\frac{3}{2}\varepsilon_{dd} \frac{\omega_x^2\kappa_x \kappa_y
\gamma^{2}}{\zeta} \beta_y\right). \label{alpha}
\end{eqnarray}
In the limit $\varepsilon_{dd}=0$ the solutions of
Eq.~(\ref{alpha}) are independent of $g$ and $\gamma$. However, for
$\varepsilon_{dd} \ne 0$ the solutions to Eq.~(\ref{alpha}) are
dependent on both the strength of the dipolar interactions and the
aspect ratio of the trap.
\begin{figure}
\centering
\includegraphics[width=8.5cm]{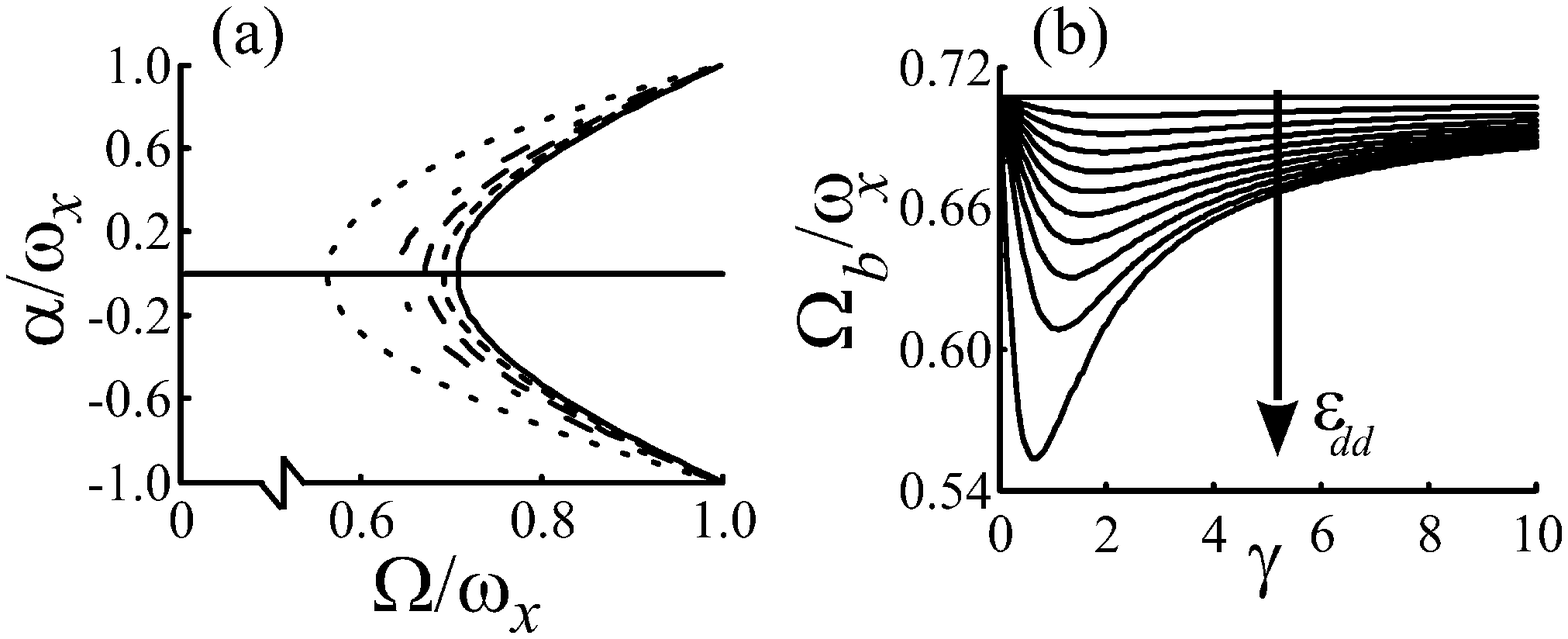}
\vspace{-8.0cm}\caption{(a) The irrotational fluid velocity,
$\alpha$, as a function of the trap rotational frequency, $\Omega$,
as obtained from Eq. ~(\ref{alpha}), for $\gamma=1$, $\epsilon=0$
and $\varepsilon_{dd}=0$ (solid curve), $\varepsilon_{dd}=0.25$
(short dashed curve), $\varepsilon_{dd}=0.5$ (long dashed curve),
$\varepsilon_{dd}=0.75$ (dash dotted curve) and
$\varepsilon_{dd}=0.99$ (dotted curve). (b) The bifurcation point
$\Omega_b$ versus $\gamma$ for different dipolar interactions
strengths; $\varepsilon_{dd}$ increases, in the direction of the
arrow, from $0$ to $0.9$ in steps of $0.1$, for the first ten
curves, the lowest curve is for $\varepsilon_{dd}=0.99$.}
 \vspace{-0.5cm}
\end{figure}

Introducing the parameter
$\epsilon=(\omega_y^2-\omega_x^2)/(\omega_x^2+\omega_y^2)$ to define
the anisotropy of the trap, we evaluate Eqs.~(\ref{kxky})
and~(\ref{alpha}) self-consistently to determine the static
solutions of the hydrodynamical equations of motion in the rotating
frame. Figure 1(a) shows the solutions to Eq.~(\ref{alpha}) for
various values of $\varepsilon_{dd}$ with $\gamma=1$  and
$\epsilon=0$.
For $\varepsilon_{dd}=0$ (solid curve)
we regain the results of Refs. \cite{Recati,Sinha} with a
bifurcation point at $\Omega_b=\omega_x/\sqrt{2}$ which exactly
coincides with the vanishing of the energy of the
quadrupolar mode in the rotating frame. For $\Omega < \Omega_b$, one
solution, corresponding to $\alpha=0$, is found. For $\Omega >
\Omega_b$, three solutions
appear, $\alpha=0$ and $\alpha=\pm \sqrt{2
\Omega^2-\omega_x^2}/\omega_x$ \cite{Recati}. The two additional
solutions are a consequence of the quadrupole mode being excited for
$\Omega \ge \omega_x/\sqrt{2}$. Actually, it is a remarkable feature
of the pure s-wave case that these solutions do not depend upon $g$.
This is because in the TF limit surface excitations with angular
momentum $\hbar l= \hbar q_{l} R$, where $R$ is the TF radius and
$q_{l}$ is the quantized wave number, obey the classical dispersion
relation $\omega_{l}^2=(q_{l}/m) \nabla_{R} V$  involving the local
harmonic potential $V=m \omega_{x}^{2}R^{2}/2$  evaluated at $R$
\cite{Pitaevskii&StringariBook}. Consequently $\omega_{l}=
\sqrt{l}\omega_{x}$, which is independent of $g$. However, in the
case of long-range dipolar interactions the potential $\Phi_{dd}$ of
Eq.\ (\ref{eq:Phidd}) gives non-local contributions, breaking the
simple dependence of the force $-\nabla V$ upon $R$ \cite{Duncnote}.
Thus, we expect the resonant condition for exciting the quadrupolar
mode, i.e. $\Omega_b=\omega_l/l$ (with $l=2$), to change with
$\varepsilon_{dd}$.
In Fig. 1(a) we see that this is the case: as dipole interactions
are introduced, our solutions change and the bifurcation point
($\Omega_b$) moves to lower frequencies.


In contrast to the s-wave case, not only the magnitude of the
dipolar coupling $\varepsilon_{dd}$, but also the  \emph{shape} of
the BEC determines the potential $\Phi_{dd}$. For an oblate
($\kappa_{x,y}>1$) BEC, more dipoles lie side-by-side, thus giving a
net repulsive interaction, in comparison to the prolate
($\kappa_{x,y} <1$) case where a majority sit end-to-end, in which
configuration the dipolar interaction is attractive.
 In the extreme
limits of $\kappa_{x,y} \rightarrow 0$ and $\kappa_{x,y} \rightarrow
\infty$ the angular dependence of the interactions plays no role and
the gas behaves conventionally, but in the intermediate regime
the role of $\kappa_{x,y}$ and hence the aspect ratio of the trap is important.
 In Fig. 1(b) we plot
$\Omega_b$ as a function of $\gamma$ for various values of
$\varepsilon_{dd}$. For $\varepsilon_{dd}=0$ we find that the
bifurcation point remains unaltered at $\Omega_b=\omega_x/\sqrt{2}$
as $\gamma=\omega_z/\omega_x$ is changed \cite{Recati,Sinha}.  As
$\varepsilon_{dd}$ is increased the value of $\gamma$ for which
$\Omega_b$ is a minimum changes from a trap shape which is oblate
($\gamma
> 1$) to prolate ($\gamma < 1$).

Consider now the effect of finite trap anisotropies  ($\epsilon
> 0$). In Fig. 2(a) we have plotted the solutions to
Eq.~(\ref{alpha}) for various values of $\varepsilon_{dd}$ with
$\gamma=1$ and $\epsilon=0.02$.
As in the case without dipolar interactions \cite{Recati,Sinha} the
solution $\alpha=0$ is no longer a solution for all $\Omega$. The
effect of introducing the anisotropy, in the absence of dipolar
interactions, is to {\em increase} the bifurcation frequency
$\Omega_b$. Turning on the dipolar interactions, as in the case of
$\epsilon=0$, {\em reduces} the bifurcation frequency.

We now analyze two procedures for generating an instability by
tracing different paths on Fig. 2(a).
\newline
{\it Procedure I:} $\Omega$ is fixed at $\Omega
> \Omega_b(\epsilon=0)$ and the trap anisotropy is adiabatically
turned on. Following analyses for conventional BECs
\cite{Recati,Rapid,Emil} we find that as $\epsilon$ is increased
adiabatically, from zero, the $\alpha=0$ solution moves to negative
values of $\alpha$ and the BEC follows this route. However, as
$\epsilon$ is increased further the edge of the lower branch
$\Omega_b(\epsilon)$ shifts to higher frequencies. At some critical
value of $\epsilon$, $\Omega_b(\epsilon) = \Omega$, the lower branch
ceases to be a solution. In this manner the evolution of the static
solutions induces instability, which in conventional BECs has been
experimentally and theoretically linked to vortex lattice formation
\cite{Madison,Rapid}. As the dipole interactions are increased the
bifurcation frequency is reduced and the range of $\Omega$ for which
this type of instability can occur increases from
$[\omega_x/\sqrt{2},\omega_x]$ to $[0.5\omega_x,\omega_x]$. In
addition, dipolar interactions increase the value of $\epsilon$ for
which lower branch solutions exist.
\newline
{\it Procedure II:} $\epsilon$ is fixed and $\Omega$ is introduced
adiabatically. In this case the BEC will follow the upper branch of
the solutions. For conventional BECs, a vortex lattice will form
\cite{Sinha,Rapid} when the upper branch solutions ($\alpha
>0 $) to Eq.~(\ref{alpha}) become dynamically unstable. Below we generalize
the analysis of Ref. \cite{Sinha} to examine the dynamical
stability of the solutions to Eq.~(\ref{alpha}).
\begin{figure}
\centering
\includegraphics[width=8.5cm]{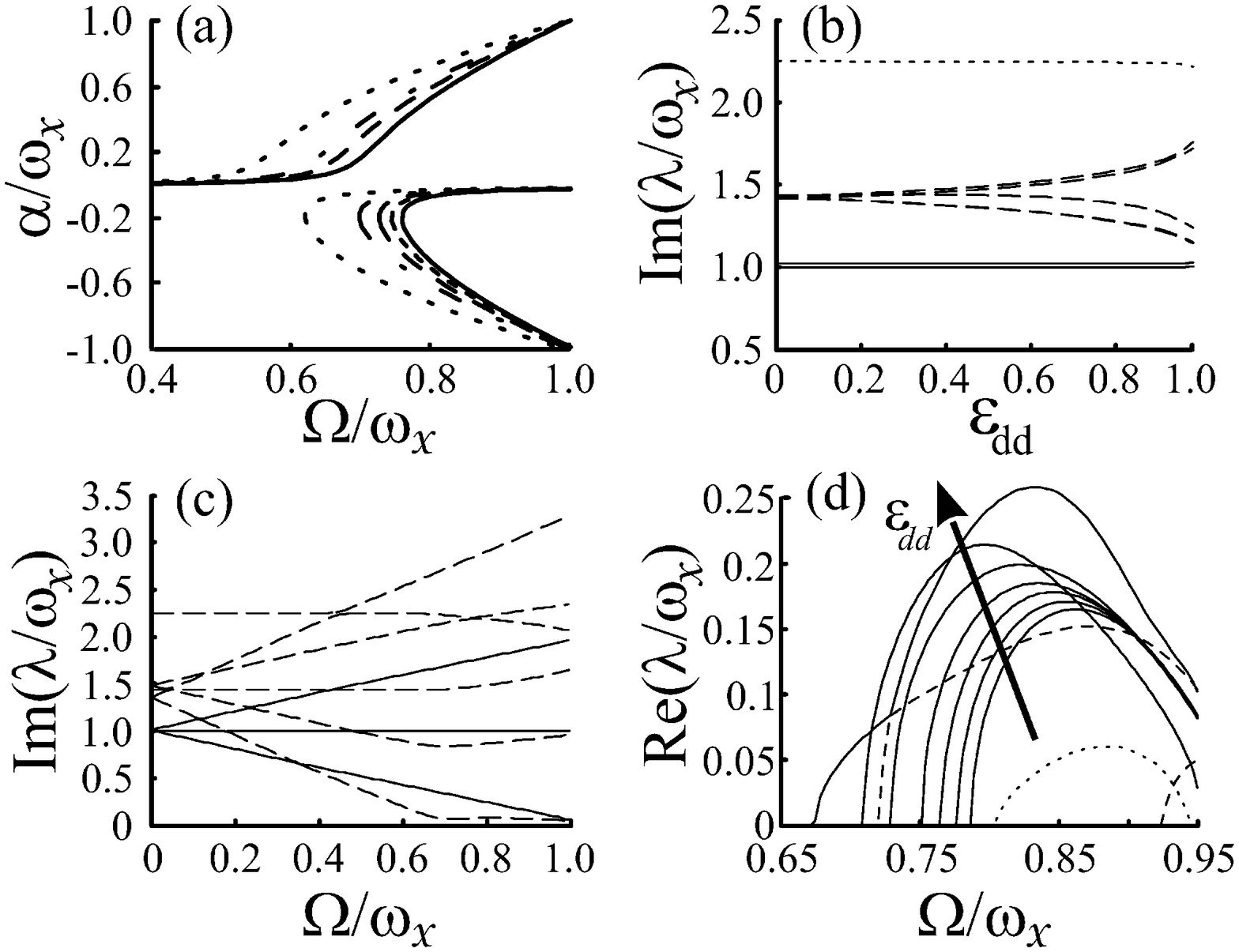}
\vspace{-5.0cm}\caption{(a) $\alpha$ vs. $\Omega$, as in Fig 1(a)
but for $\epsilon=0.02$. (b) The positive imaginary eigenvalues of
Eq. (15) as a function of $\varepsilon_{dd}$ for $\Omega=0$,
$\epsilon=0.02$, $\gamma=1$ and $n=2$. The three solid curves [two
at Im$(\lambda)=\omega_x$ and one at
Im$(\lambda)=\omega_x(1+\epsilon)^{0.5}=\omega_y$] are the
frequencies of the center of mass modes of the BEC. The dashed
curves are the frequencies of the breathing modes of the BEC. (c)
The positive imaginary eigenvalues of Eq. (15), as a function of
$\Omega$ for $\varepsilon_{dd}=0.5$, $\epsilon=0.02$, $\gamma=1$ and
$n=2$. The solid (dashed) curves are the frequencies of the center
of mass (breathing) modes of the BEC.
 (d) The maximum positive real eigenvalues of Eq. (15) (solid curves), as a function of $\Omega$,
 for $\epsilon=0.02$,
$\gamma=1$, $n=3$ and $\varepsilon_{dd}=0,0.2,0.4,0.6,0.8,0.95$ and
$0.98$; $\varepsilon_{dd}$ increases in the direction of the arrow.
The short and long dashed curves are additional positive eigenvalue
solutions for $\varepsilon_{dd}=0.95$ and $0.98$ respectively.}
 \vspace{-0.5cm}
\end{figure}

Consider small perturbations in the BEC density and phase of the
form $\rho=\rho_0+\delta \rho$ and $S=S_0+\delta S$ then, via
Eqs.~(\ref{continuity}, \ref{hydro}) the dynamics of such
perturbations can be described, to first order, as
 \begin{eqnarray}
\frac{\partial }{\partial t} \left[\begin{array}{c}
 \delta S \\
  \delta \rho \\
\end{array}
\right] = -\left[\begin{array}{cc}
 \underline{v}_c \cdot \nabla & \frac{g}{m}\left(1+\varepsilon_{dd}K\right) \\
 \nabla \cdot \rho_0
\nabla & \left[\left(\nabla \cdot
\underline{v}\right)+\underline{v}_c \cdot \nabla \right]  \\
\end{array}
\right] \left[\begin{array}{c}
 \delta S \\
  \delta \rho \\
\end{array}
\right]
 \end{eqnarray}
where $K=-3\frac{\partial^2}{\partial z^2}\int dxdydz/(4 \pi \left|
\underline{r}^\prime-\underline{r} \right|)-1$ and
$\underline{v}_c=\underline{v}-\underline{\Omega}\times\underline{r}$.
As in Ref. \cite{Sinha} we consider a polynomial ansatz, of order
$n$ in the coordinates $x,y,z$ and evaluate the evolution operator
for the perturbations. If one or more of the eigenvalues, $\lambda$,
has a positive real component the stationary solution is dynamically
unstable. However, imaginary eigenvalues correspond to stable
oscillatory modes of the system \cite{Castin2}. Below we consider
both the stable and unstable modes of the upper branch static
solutions, for $\gamma=1$ and $\epsilon=0.02$.

In Fig. 2(b) we plot the positive imaginary eigenvalues of Eq. (15)
for $n=2$ as a function of the dipolar interaction strength with
$\Omega=0$. As expected we find three modes associated with center
of mass oscillations [solid curves in Fig. 2(b)], two at
Im$(\lambda)=w_x=w_z$ and one at Im$(\lambda)=w_y$. These modes are
independent of the strength of the dipolar interactions since they
do not alter the shape of the BEC.  The higher frequency modes
(dashed curves) are associated with the breathing modes of the
system \cite{Castin2}. Since these modes do alter the shape of the
BEC and the resulting dipolar interactions we find that they are
dependent upon $\varepsilon_{dd}$. However, the breathing mode at
Im$(\lambda)=\omega_x \sqrt{5}$ is associated with a perturbation
 in $x$, $y$ and $z$ which is
equivalent to a  uniform re-scaling of the density and as such the
frequency of this mode is almost independent of the dipolar
interaction strength, as can be seen in Fig. 2(b). Fig. 2(c) shows
how these modes shift as a function of $\Omega$, for
$\varepsilon_{dd}=0.5$. Under the rotation the motion in $x$ and $y$
is coupled. Thus the frequency of the center of mass modes (solid
curves) is shifted in the $x-y$ plane
\cite{Pitaevskii&StringariBook}
whilst remaining constant, at $\omega_z$, in the $z$ direction.
Again these modes are independent of the strength of the dipolar
interactions. Conversely, the breathing modes (dashed curves) have a
relatively complex dependence on $\varepsilon_{dd}$ and the rotation
frequency.

Finally we consider the real positive eigenvalues of Eq. (15),
associated with regions of instability for the upper branch static
solutions. In the limit of $\varepsilon_{dd}=0$ we reproduce Fig. 2
of Ref. \cite{Sinha}, with the solutions being unstable in the range
$[0.78 \omega_x,\omega_x]$ for $\epsilon=0.02$. In Fig. 2(d) we have
plotted the real positive eigenvalues, Re$(\lambda)$, of Eq. (15),
as a function of $\Omega$ for various values of $\varepsilon_{dd}$
with $n=3$. For higher values of $\varepsilon_{dd}$ [$0.95$ and 0.98
in Fig. 2(d)] there can be more than one real positive eigenvalue,
thus we define the region of instability as the range over which
max[Re$(\lambda)>0$], as shown by the solid curves in Fig. 2(d)
\cite{note_n}. As the dipolar interaction strength is increased the
lower bound in $\Omega$ for the unstable region is reduced. For
example, for $\varepsilon_{dd}=0.6$ the range of rotation
frequencies where the upper branch solution is unstable is $[0.75
\omega_x, \omega_x]$, this increases to $[0.67 \omega_x, \omega_x]$
for $\varepsilon_{dd}=0.98$.

By calculating the static hydrodynamic solutions of a rotating
dipolar BEC and studying their dynamical stability, we have
predicted the regimes of instability of the condensate.
In general we find that the bifurcation frequency, $\Omega_b$,
decreases in the presence of dipolar interactions. Thus, for a fixed
$\Omega$ [at $\Omega > \Omega_b(\epsilon =0)$] and an adiabatic
increase in $\epsilon$, the critical anisotropy at which we expect
instability to occur will be higher than for a conventional BEC.
Furthermore, we find that the size of this shift not only depends on
the strength of the dipolar interactions but also on the aspect
ratio of the trap, with the maximal shift being for $\gamma < 1$
($\epsilon_{dd} \rightarrow 1$). For a fixed anisotropy and an
adiabatic increase in $\Omega$ we find that as $\varepsilon_{dd}$ is
increased the lower bound on the rotation frequency at which a
rotating dipolar gas will be unstable to perturbations is decreased.
In conventional BECs these instabilities have been related to vortex
lattice formation \cite{Rapid}. This occurs, primarily, because the
instability disrupts the BEC at an $\Omega$  which is greater than
the rotation frequency at which it is energetically favorable to
have a vortex state \cite{Lundh97}. However, in a prolate trap the
rotational frequency at which it is energetically favorable to form
a vortex in a dipolar BEC grows rapidly as  $\varepsilon_{dd}$ is
increased \cite{ODell06} and can exceed the frequency at which we
expect an instability to occur. The final state under these
circumstances warrants further investigation.

We thank Subhasis Sinha and Yvan Castin  for helpful discussions
and acknowledge the financial support of the ARC, the EPSRC, the
University of Melbourne and QIP IRC (GR/S82176/01).


\begin{thebibliography}{99}
\bibitem{note1} The limit where the zero-point kinetic  energy is negligible compared to the potential and interaction energies.
\bibitem{note_n} For $n>3$ we find that although there are more eigenvalues
the region of instability, defined by max[Re$(\lambda)>0$], remains
the same.
\end{thebibliography}
\end{document}